\documentstyle[pre,multicol,aps]{revtex}
\newcommand{\BEQ}{\begin{equation}}
\newcommand{\EEQ}{\end{equation}}
\newcommand{\BEA}{\begin{eqnarray}}
\newcommand{\EEA}{\end{eqnarray}}
\newcommand{\nn}{\nonumber \\}
\renewcommand{\d}{{\rm d}}
\newcommand{\oz}{{(0)}}
\newcommand{\Tf}{T_0} 
\newcommand{\Tg}{T_g}
\newcommand{\Vn}{V_0}
\newcommand{\Tfin}{T_0} 

\newcommand{\half}{\frac{1}{2}}
\begin{document} 
\draft
\title
{Optimizing the classical heat engine.}
\author{A.E. Allahverdyan$^{1,3)}$ and Th.M. Nieuwenhuizen$^{2)}$}
\address{$^{1)}$ Institute for Theoretical Physics, 
$^{2)}$ Department of Physics and Astronomy,\\ 
University of Amsterdam,
Valckenierstraat 65, 1018 XE Amsterdam, The Netherlands 
\\ $^{3)}$Yerevan Physics Institute,
Alikhanian Brothers St. 2, Yerevan 375036, Armenia }

\maketitle

\begin{abstract}
A pair of systems at different temperatures is a classic environment for 
a heat engine, which produces work during the relaxation to 
a common equilibrium. It is generally believed that a direct
interaction between the two systems will always 
decrease the amount of the obtainable work, due to 
inevitable dissipation. Here a situation is reported where, 
in some time window, work can be gained due the direct coupling, 
while dissipation is relevant only for much larger times.
Thus, the amount of extracted work {\it increases}, 
at the cost of a change of the final state.
\end{abstract}


\begin{multicols}{2}
{\it Introduction.}|
The classical problem of thermodynamics is the determination of 
the maximal amount of work that can be extracted from a non-equilibrium 
system, during the relaxation to the equilibrium state 
\cite{callen,landau,ma,a}. 
In recent years the interest in this time-honored problem 
was renewed (see \cite{a} for review). 
In many cases the limits proposed by the founders of the classic 
thermodynamics have appeared as too idealistic, 
and attention was focused on the study of dissipative effects, 
which restrict the abilities of realistic heat engines \cite{a}.
One of the most popular examples of that kind is a direct
interaction between the thermal baths 
that drive the standard heat engine \cite{aa}. 
Evidently, this is a way to dissipate energy. 
The common opinion, expressed in textbooks \cite{callen,landau,ma}
is that dissipation is the main, if not the only effect of any direct 
interaction. 

The purpose of the present paper is to show that
a direct interaction between the baths may have energy transfer, 
rather than energy dissipation, as its main physical effect.
The reason is that the relevant timescale of the transfer 
arising from the direct coupling can be widely separated from the 
dissipative timescale.
This leads to an optimization of the classical heat engine, 
in a certain time window. 
As a result, the final equilibrium state of 
the heat baths is changed.
The mechanism arises from the recently proposed 
steady adiabatic state \cite{allah} (see also 
\cite{theohammer,theolongtermo} in the related context). 

First, we shall reproduce the classical discussion of the maximal
amount of work that can be extracted from a non-equilibrium system.
The underlying ideas are, of course, well-known, and we shall need 
them for a careful interpretation of our main result. 

{\it The classical analysis.}| 
Consider two 
subsystems which have different temperatures $T_1$ and $T_2$. Depending 
on the concrete relaxation process, the whole thermally isolated system 
will produce different amounts of work. This model is
well-known and described in textbooks on thermodynamics 
\cite{callen,landau,ma}.
We shall assume that the total volume of the 
system is unchanged by the relaxation
(although it can vary during the process),
since we wish to take into account only the work that can be done 
due to the non-equilibrium initial state, and 
not due to the general expansion. Denoting by $U_i$, $S_i$ 
the initial energy and entropy of the system,
we get the following expression for the work $W$ 
performed by the system
\begin{equation}
\label{1}
W=U_i-U(S),
\end{equation}
where $U(S)$ is energy of the final equilibrium state as 
a function of its entropy. 
Because the temperature is positive, $U$ is a monotoneous function of $S$
($\partial U/\partial S|_V=T>0$ in equilibrium).
Therefore, $W$ is maximal when $S$ is as small as possible. 
Since the whole system is thermally isolated, 
the second law demands $S\geq S_i$. 
The maximal amount of work is attained for $S=S_i$,
i.e., for a {\it reversible} process towards equilibrium. 
It is believed that for obtaining the maximal amount of
 work any direct interaction
between subsystems should be removed, because it would induce 
{\it irreversible} relaxation.  Thus, a third 
body (``engine'') should operate between our subsystems 
to perform the work. At the end of the relaxation process 
the working body must return to its initial state.
The expression of the total amount $W$ of the extracted work 
can be given as 
\begin{equation}
\label{2}
W=U_1(T_1)+U_2(T_2)-U_1(\Tfin)-U_2(\Tfin),
\end{equation}
where the final temperature $\Tfin$ is determined by
the reversibility condition
\begin{equation}
\label{3}
S_1(T_1)+S_2(T_2)=S_1(\Tfin)+S_2(\Tfin),
\end{equation}
and $U_k$, $S_k$ ($k=1,2$) are 
energy and entropy of the corresponding subsystems. 
The reversible relaxation process consists of an
infinite amount of elementary cycles of the third body. 
The famous Carnot process is the most 
representative example for such a cycle. The local efficiency $\eta $
of the optimal cycle, defined as the maximal extracted work during the 
cycle divided by the input energy, depends {\it only}
on the conditions of reversibility and conservation of energy:
$\eta =1- (T_1/T_2)$, if $T_1<T_2$ \cite{callen,landau,ma}.
This efficiency is universal and system-independent, emphasizing
the power of thermodynamics. In contrast, the
maximal amount of the extracted work (\ref{2}) is not universal, and
can vary from one system to another.

{\it The steady adiabatic state.}|
Steady adiabatic systems have two distinctive properties \cite{allah}:
1) different subsystems are not in the mutual equilibrium, but 
possess different temperatures, 2) the local characteristic 
relaxation times of these subsystems are very 
well separated. 
Let us consider a statistical system which has two subsystems 
with coordinates $x_1$ and $x_2$. 
($x_1$ and $x_2$ can thus also code also a set of variables; we
will not denote that explicitly.) 

The corresponding relaxation times are
denoted by $\Gamma _1$ and $\Gamma _2$. The  condition of well
separated timescales is ensured by
$\gamma =\Gamma _1/\Gamma _2\ll 1$.
The stationary distribution of this system can be found 
from the following heuristic arguments (for a more rigorous 
presentation see \cite{allah}). 
Due to the large difference between relaxation times the 
$x_1$-subsystem comes to an equilibrium while the 
$x_2$-variable is almost unchanged. The corresponding
Gibbs stationary distribution  reads
\begin{equation}
\label{5}
P(x_1|x_2)=\frac{1}{Z(x_2)}\exp [-\beta _1 H(x_1, x_2)],
\end{equation}
where 
\begin{equation}
\label{hamham}
H(x_1,x_2)=H_1(x_1)+H_2(x_2)+gH_{int}(x_1,x_2)
\end{equation}
is the total system's Hamiltonian, $T_1=1/\beta _1$ is 
the temperature of the $x_1$-subsystem, and $Z(x_2)$ is the 
partition sum at fixed $x_2$. 
For obtaining the coarse-grained distribution $P(x_2)$ 
notice that after integrating out of 
the vast variable
$x_1$, the effective Hamiltonian for the slow variable $x_2$ is 
just $-T_1\ln Z(x_2)$, the free energy of the 
$x_1$-subsystem at fixed $x_2$. 
The stationary distribution of $x_2$ then reads
\begin{equation}
\label{6}
P(x_2)=\frac{Z^{T_1/T_2}(x_2)}{{\cal Z}}, \qquad
{\cal Z}=\int {\rm d}x_2 Z^{T_1/T_2}(x_2).
\end{equation}
The complete stationary distribution 
can now be written as $P(x_1,x_2)=P(x_2)P(x_1|x_2)$. 
The mean energy and entropy of the system 
are given by the following general definitions
\begin{eqnarray}
\label{7}
&&U=\int {\rm d}x_1{\rm d}x_2 H (x_1,x_2)P(x_1,x_2),\\
\label{8}
&&S=-\int {\rm d}x_1{\rm d}x_2 P(x_1,x_2)\ln P(x_1,x_2),
\end{eqnarray}
$S$ can be decomposed as the sum of the entropies of 
the slow and fast subsystems, $S=S_1+S_2$, with
\begin{eqnarray}
\label{9}
&& S_1=\int{\rm d} x_2\,P(x_2)[-\int {\rm d}x_1 P(x_1|x_2)\ln P(x_1|x_2)], 
\nonumber \\
&&S_2=-\int {\rm d}x_2 P(x_2)\ln P(x_2). 
\end{eqnarray}
$S_1$ is the entropy of the fast variable $x_1$ at 
$x_2$, averaged $x_2$,
while $S_2$ is the entropy of this slow variable itself. 
As was shown in \cite{allah} the considered system admits
a {\it thermodynamical} description. In particular,
defining the free energy as $F= -T_2\ln {\cal Z}$, we get:
\begin{equation}
\label{u1}
F=U-T_1 S_1-T_2 S_2,
\end{equation}
where the entropies can also be obtained 
by the standard relations: 
\begin{equation}
\label{entropy}
S_1=-\left. \frac{\partial F}{\partial T_1 }\right \vert_{T_2}, \qquad
S_2=-\left. \frac{\partial F}{\partial T_2} \right \vert_{T_1}. 
\end{equation}
The energy and entropies are constant in the steady state, 
but a direct coupling induces a steady entropy 
production at rate $\dot S$ and an energy dissipation at rate $\dot{\Pi}$. 
These quantities were analyzed in Ref.
\cite{allah}, and the obtained formulas read 
$\dot{\Pi}=T_2\dot S+{\cal O}(\gamma^2)$, with 
\begin{eqnarray}
\label{s1}
&&\dot{{S}} = \gamma g^2\frac{(T_1-T_2)^2 }{\Gamma _1\,T_1^2T_2}
\int {\rm d}x_2\,{\rm d}x_1\,P(x_1,x_2)\times\nn
&& \left[\frac{\partial H_{int}(x_1,x_2)}{\partial x_2} -
\int {\rm d} y P(y |x_2)\frac{\partial H_{int}(y,x_2)}
{\partial x_2} \right ]^2+{\cal O}(\gamma^2)\nn
&&
\end{eqnarray}
Although these results were obtained for the strictly 
steady state, they can be applied also in the time-dependent 
case, if the characteristic time of this {\it quasi-stationary}
process is much larger than the largest relaxation time,
$\Gamma _2$. For instance, to obtain the time-dependent
distribution function for the case of slowly (adiabatically) 
changing temperatures or other parameters, 
one just inserts these time-dependent 
values directly in Eqs.~(\ref{5}, \ref{6}).
In this context, the change of free energy (\ref{u1})
can be shown to be the adiabatic work ${\cal W}_{ad}$ 
done on the system, when varying a parameter
$\alpha $ (for example, the width of the potential, or a coupling 
constant) from the initial value $\alpha _i$ to final value $\alpha _f$
at constant temperatures:
\begin{eqnarray}
\label{work}
{\cal W}_{ad}~&&=\int _{\alpha _i}^{\alpha _f}\d \alpha 
\int\d x_1\d x_2P(x_1,x_2,\alpha )
\frac{\partial H(x_1,x_2,\alpha)}{\partial \alpha}\nonumber \\
&&=F(T_1,T_2;\alpha _f)-F(T_1,T_2;\alpha _i)
\end{eqnarray}
This is fully analogous to the property of the usual 
(single-temperature) free energy.

The dissipative effects (given by Eq.~(\ref{s1})) 
are small for small $\gamma$, $g$. 
Let us neglect them for the moment; later we shall 
show that this is allowed in a certain timewindow.

{\it The maximal amount of work extracted from 
the steady adiabatic state.}|
Certainly, we can apply the above-mentioned general 
analysis, concerning the maximal work,
to our adiabatic system.
In fact, this analysis does not use any concrete 
property of the initial non-equilibrium state, but 
we should take into account that 
our subsystems interact directly, and not only 
through the third body. 
As compared to the case without direct coupling,
the system will now relax to a different equilibrium state, 
and this is the reason why one can get more work done by it.
The total amount of the 
gained work can be again written as 
\begin{equation}
\label{10}
W(g)=U(T_1, T_2)-U(\Tg,\Tg),
\end{equation}
and the temperature $\Tg$ of the final equilibrium 
state is defined from the condition of reversibility
\begin{equation}
\label{11}
S(T_1, T_2)=S(\Tg,\Tg).
\end{equation}
This condition involves the total entropy of the interacting
subsystems, and now assumes that there are no additional 
sources of dissipation besides (\ref{s1}).

We shall investigate Eqs.~(\ref{10}, \ref{11})
to first order in the small parameter $g$. 
Hereafter quantities of the  order  $g^0$ and $g^1$ 
will be indicated by the index $0$ and $1$, respectively. 
It is evident from (\ref{hamham}) that to order $g^1$ it holds 
\begin{equation}
\label{free}
F=F^{(0)}+g\Vn \EEQ
where
\BEQ  \Vn(T_1,T_2)=\int \d x_1\d x_2P(x_1,x_2) H_{int}(x_1,x_2).
\end{equation}
Using 
Eqs.~(\ref{7}-\ref{entropy}), one gets
\begin{eqnarray}
\label{do}
&&S_1 =S_1^\oz+gS_1^{(1)}=S_1^\oz-g\partial _{T_1}\Vn ,\\
&&S_2 =S_2^\oz+gS_2^{(1)}=S_2^\oz-g\partial _{T_2}\Vn ,\\
&&S =S^\oz+gS^{(1)}=S^\oz-g\{ \partial _{T_1}+\partial _{T_2} \}\Vn ,\\
&&U=U^\oz+gU^{(1)}=U^\oz+g\{ 1-T_1\partial _{T_1}-T_2\partial _{T_2} \}
\Vn \label{ut}
\end{eqnarray}
Let us now obtain from Eq. (\ref{11}) an expression for the final 
temperature $\Tg$ to order $g$,  given the value of $\Tf$, 
the final temperature for $g=0$,
\begin{equation}
\label{Tf+g}
\Tg=\Tf\left (1+g\frac{S^{(1)}(T_1,T_2)
-S^{(1)}(\Tf,\Tf) }{C_1+C_2}\right ).
\end{equation}
Here $C_k=T_k\,\partial S^{(0)}_{k}/\partial T_k|_V=
\partial U^{(0)}_{k}/\partial T_k|_V$, with $k=1,2$, are
the heat capacities of the subsystems when they are uncoupled.
Starting from Eq.~(\ref{10}) and using Eqs. (\ref{do})-(\ref{ut}),
we finally derive the excess work at order $g$
\begin{eqnarray}
\label{re}
&&W^{(1)}=\lim_{g\to 0}\frac{W(g)-W(0)}{g}\\
&&= \Vn(T_1,T_2)-
\Vn(\Tf,\Tf)+\sum_{k=1}^2(\Tf-T_k)\partial _{T_k}\Vn(T_1,T_2).\nonumber
\end{eqnarray}
This is the first main result of our work. It remains to be shown in a
specific example that this quantity can be positive.
Let us first point out that 
further simplifications occur when $T_1$ is close to $T_2$. 
To first order in the parameter $T_1-T_2$ eq. (3) gives us 
\begin{equation}
\label{\Tf}
\Tf=\frac{C_1+C_2}{C_1T_1+ C_2T_2}T_1T_2,
\end{equation}
while Eq. (\ref{re}) can be approximated by
\begin{equation}
\label{mi}
W^{(1)}=\sum_{k=1}^2(\Tf-T_k)(\partial _{T_k}
\Vn-\{ \partial _{T_k}\Vn\}|_{T_1=T_2=\Tf}).
\end{equation}

To illustrate the general results let us present
a concrete model, where
the direct interaction increases the total amount of work:
$W(g) > W(0)$. One of the most popular models of the thermal bath
is a set of harmonic oscillators \cite{gardiner},
which is frequently used to derive kinetic equations or to
gain fundamental insight. Following this well-established tradition,
we shall model our first (second) thermal bath by $N_1$$ (N_2$)
oscillators at temperature $T_1$ ($T_2$), and weakly-anharmonic 
interaction:
\begin{equation}
\label{ham}
H=\frac{1}{2}\sum_{i=1}^{N_1}x_{1,i}^2+
\frac{1}{2}\sum_{i=1}^{N_2}x_{2,i}^2
+g\sum_{i=1}^{N}x_{1,i}^2x_{2,i}^2, 
\end{equation}
where $g>0$, and $N_1, N_2\ge N$. 
It is straightforward to show that to order $g$ one has the partial
partition sum
\BEQ Z(x_2)=T_1^{N_1/2}\,
\exp\left(-\frac{1}{2}\beta_1\sum_{i=1}^{N_2}x_{2,i}^2-
g\sum_{i=1}^N x_{2,i}^2\right)\EEQ
and the full partition sum
\BEQ {\cal Z}=T_1^{\frac{1}{2}N_1T_1/T_2}\,T_2^{\frac{1}{2}N_2}
(1+2gT_1)^{-\frac{1}{2}N}\EEQ
The latter result yields the free energy $ F=-T_2\ln {\cal Z}$
\BEQ 
\label{dod1}
F=-\half N_1T_1\ln T_1-\half N_2T_2\ln T_2+gNT_1T_2\EEQ
in agreement with the fact that 
$\Vn=
NT_1T_2$. 
According to previous rules we derive
\BEA\label{S1m=}
 S_1&=&\frac{1}{2}N_1(\ln T_1+1)-gNT_2\\
 S_2&=&\frac{1}{2}N_2(\ln T_2+1)-gNT_1
\label{S2m=}\EEA
The internal energy follows as $U=F+T_1S_1+T_2S_2$,
\BEQ 
\label{w1}
U=\half N_1T_1+\half N_2T_2-gNT_1T_2\EEQ
Notice that the sign of the order $g$ correction is negative,
due to entropic effects.
To determine $\Tf$ from 
Eq.~(\ref{3}) we need entropies $S^{(0)}_1$, $S^{(0)}_2$,
that can be read of from Eqs. (\ref{S1m=}), (\ref{S2m=}) at $g=0$. 
One obtains $\Tf=T_1^{\nu_1}T_2^{\nu_2}$ and then from Eq.~(\ref{Tf+g})
\begin{eqnarray}
\label{12}
\Tg=T_1^{\nu_1}T_2^{\nu_2}[1+2g\nu(2T_1^{\nu_1}T_2^{\nu_2}-T_1-T_2)],
\end{eqnarray}
where $\nu =N/(N_1+N_2)$, $\nu_k=N_k/(N_1+N_2)$, $k=1,2$. 
According to Eq.~(\ref{1}) and using
$U_k^{(0)}=N_kT_k/2$, one gets for $g=0$ the maximal amount of work 
\begin{equation}
\label{zeka}
W(0)=\frac{1}{2} N_1(T_1-T_1^{\nu_1}T_2^{\nu_2})
+\frac{1}{2} N_2(T_2-T_1^{\nu_1}T_2^{\nu_2})
\end{equation}
Taking into account that $\Vn(T_1,T_2)=NT_1T_2$,
one gets from Eq.~(\ref{re}) a non-negative
shift in the maximal of amount of work
\begin{eqnarray}
\label{13}
gW^{(1)}
&=& 
gN(T_1^{\nu_1}T_2^{\nu_2}-T_1)(T_2-T_1^{\nu_1}T_2^{\nu_2})\ge 0.
\end{eqnarray}
The equality is realized in the trivial case $T_1=T_2$.
Only  for $g<0$ this mechanism would reduce the work.


Eq.~(\ref{re}), 
and especially our model-dependent result
(\ref{13}), show that the full amount of the extracted work can
{\it increase} due the direct coupling.
Let us now return to the dissipative effects.
For the considered model the energy dissipated per unit of
time $\dot \Pi=T_2\dot S$ can be derived from Eq. (\ref{s1}). 
It reads \cite{allah} 
\begin{equation}
\label{16}
\dot{ \Pi }=\frac{8g^2N}{\Gamma _1}\gamma T_2(T_1-T_2)^2+
{\cal O}(\gamma ^2)
\end{equation}
Our aim is now to obtain the characteristic time ${\cal T}$,
after which the 
energy dissipated due to the direct coupling is comparable
with the energy $gW^{(1)}$ gained according to Eq.~(\ref{13}). 
For the dissipated energy an upper estimate can be given as
$\dot\Pi\,{\cal T}$, and we get from 
$\dot\Pi\,{\cal T}=gW^{(1)}$, 
\begin{equation}
\label{time}
{\cal T}= 
\frac{\Gamma _2}{8g}
\,\frac{(T_1^{\nu_1}T_2^{\nu_2}-T_1)(T_2-T_1^{\nu_1}T_2^{\nu_2})}
{(T_1-T_2)^2}.
\end{equation}
To be able to neglect the dissipated energy, the duration
of our process $t$ must be much smaller than ${\cal T}$.
On the other hand, since we are getting the work in the 
relaxation process, its duration $t$ must be much higher than the
largest relaxation time $\Gamma _2$. Thus, for times
\begin{equation}
\label{t<<t}
\Gamma_{2}\ll t\ll {\cal T}
\end{equation}
it is possible to perform more work due to the presence
of the direct coupling.
The necessary condition $\Gamma_{2}\ll{\cal T}$
is realized mainly when $g$ is small.
If one was not able to complete the relaxation in the time-window
(\ref{t<<t}), then for $t\sim {\cal T}$ the gained work will be equal
to that without any direct coupling. 
The same analysis can be applied for the general case.

So far we have compared the efficiencies of two systems 
with the same initial temperatures $T_1$, $T_2$, and different values of
$g$. 
One can also compare cases of identical initial energies,
$(T_1,T_2,g=0)$ and $(\bar{T_1},\bar{T_2},g>0)$, where the
temperatures $\bar{T_1},\bar{T_2}$ are defined by 
$U(T_1,T_2,g=0)=U(\bar{T_1},\bar{T_2},g>0)$. The analysis is very similar
to 
that given above, and indicates that our main result remains valid
also in this case. There are examples of $(\bar{T_1},\bar{T_2})$, for
which  the direct coupling enhances the work.

In the context of our main result it is useful to investigate which amount 
of work ${\cal W}(0\to g)$ should be spent by external sources to switch
on 
the small coupling $g$, starting from the state with $g=0$. We shall 
consider the two extremes, very slow and very fast switching, 
which happen to give the same answer for small $g$. 
In the first case one uses 
Eqs.~(\ref{work}, \ref{free}) to obtain ($T_1$, $T_2$ are constant)
\begin{equation}
\label{sw}
{\cal W}_{ad}(0\to g)=gV_0
\end{equation}
For the very fast switching the initial state does not change, and 
the main change comes from the Hamiltonian (\ref{hamham}):
${\cal W}_{fast}(0\to g)=\langle H(g)-H(0)\rangle _0$, which for small $g$
leads to the same result as in Eq.~(\ref{sw}). Using Eqs~(\ref{re}, 
\ref{13}, \ref{sw}) one readily notices that there are temperatures,
for which ${\cal W}_{ad}(0\to g)<gW^{(1)}$, implying that the
cost for the switching is less than the gain due to coupling:
$W(0)+{\cal W}_{ad}(0\to g)<W(g)$.

{\it Conclusion.}|
Untill now it was believed that the presence of a direct interaction 
between the baths of a heat engine reduces its efficiency
\cite{callen,landau,ma,a,aa,motors}. The purpose of the present paper
is to demonstrate that it can enhance the efficiency.
Having changed the initial and final states,
a direct coupling introduces, of course, both a change in work and
dissipation. We show that the characteristic times of these two quantities
can be well-separated. For times in the window (\ref{t<<t}) the
work can be enhanced, 
though the dissipation is not yet relevant.
This additional amount of work, which can be obtained from Eqs.~(\ref{re},
\ref{mi}, \ref{13}), is provided by the modified final state
of the baths.

Finally, we will briefly discuss related studies.
Refs.~\cite{motors,sokolov} consider the local thermodynamic
efficiency of brownian motors and related models. The statement 
of this problem differs from the one considered by us, but it is 
interesting to mention that the role of a direct interaction 
between baths was studied also in this context \cite{sokolov}.

\end{multicols}
\end{document}